\documentclass[a4paper, onecolumn, 11pt]{quantumarticle}
\pdfoutput=1
\usepackage[utf8]{inputenc}
\usepackage[english]{babel}
\usepackage[T1]{fontenc}
\usepackage{amsmath}
\usepackage{hyperref}

\usepackage{tikz}
\usepackage{lipsum}

\usepackage{graphicx}
\usepackage{amsmath}
\usepackage{subcaption}
\usepackage{parskip}

\usepackage[numbers]{natbib}

\begin{document}
	
	\title{Curvature effects on stimulated parametric down-conversion process: an analog model}
	
	\author{S. Akbari-Kourbolagh}
	\affiliation{Department of Physics, University of Isfahan, Hezar Jerib, 81746-73441, Isfahan, Iran}
	\orcid{0009-0002-8032-9465}
	\email{ saeed\_akbari@sci.ui.ac.ir}
	
	\author{A. Mahdifar}
	\email{ali.mahdifar@sci.ui.ac.ir}
	\orcid{0000-0001-5152-7791}
	\affiliation{Department of Physics, University of Isfahan, Hezar Jerib, 81746-73441, Isfahan, Iran}
	\affiliation{Quantum Optics Group, Department of Physics, University of Isfahan, Hezar Jerib, 81746-73441, Isfahan, Iran}
	
	\author{H. Mohammadi}
	\email{hr.mohammadi@sci.ui.ac.ir}
	\orcid{0000-0001-7046-3818}
	\affiliation{Department of Physics, University of Isfahan, Hezar Jerib, 81746-73441, Isfahan, Iran}
	\affiliation{Quantum Optics Group, Department of Physics, University of Isfahan, Hezar Jerib, 81746-73441, Isfahan, Iran}
	
	\maketitle
	
\begin{abstract}
	In this paper, we utilize an analog model of the general relativity to investigate the influence of spatial curvature on  quantum properties of stimulated parametric down-conversion process. For this purpose, we use two-mode sphere coherent state as the input beams of the aforementioned process. These states are realization of coherent states of two-dimensional harmonic oscillator, which lies on a two-dimensional sphere. We calculate the entanglement of output states of stimulated parametric down conversion process, measured by linear entropy, and show that it depends on the spatial curvature. Furthermore, by preparing the suitable two-mode sphere coherent states, it is possible to control the entanglement between the output states in the laboratory. In addition, we consider mean number and Mandel parameter of the output states of the process and also, their cross-correlation function, as the convince measures of non-classical behaviors.
\end{abstract}

\section{Introduction}
As established by the theory of the general relativity, the presence of massive celestial bodies alters the geometric nature of the spacetime, consequently affecting its curvature. While these effects are profound on cosmological scales, their influence at the laboratory scale is weak and challenging to detect \cite{Ref1}. This is where the concept of analog models becomes significant. Numerous physical systems have been introduced to explore analogies to the effects of the general relativity, with optical systems achieving notable success among them \cite{Ref2}. For instance, the authors of Ref.\cite{Ref3} proposed an analog model for the general relativity in Rindler space, based on multilayer films. In Refs.\cite{Ref4} and \cite{Ref5}, the behavior of light near a black hole have been simulated by using metamaterials due to their engineered electromagnetic properties. Additionally, a moving dielectric medium has been employed to model light propagating in an effective gravitational field Ref.\cite{Ref6}.

Furthermore, another class of analog models for general relativity can be employed: two-dimensional curved surfaces. By considering a constant time and extracting the equatorial slice of Friedman-Robertson-Walker spacetime \cite{Ref7}, we encounter a two-dimensional curved surface with constant curvature. In these analog models, curved space is created by directly engineering the geometry of the space itself, allowing us to investigate its effects on various physical systems.

There are two approaches for examining the effects of the space curvature on quantum systems. The first involves identifying suitable quantum states into which the curvature effects can be incorporated. The second approach focuses on selecting appropriate operators whose influence on the quantum system reflects the spatial curvature being studied.

It is well-known that the set of coherent states (CSs) associated with a harmonic oscillator  \cite{Ref8} inherit the geometric properties of the space in which the oscillator is defined \cite{Ref8, Ref9}. This geometric inheritance makes CSs ideal candidates for incorporating spatial curvature in analog models of the general relativity. A particularly interesting class of coherent states, recently introduced, are sphere coherent states (SCSs), which correspond to a two-dimensional harmonic oscillator on the surface of a sphere \cite{Ref10}. These curvature-dependent CSs provide an ideal platform for investigating the effects of spatial curvature on physical phenomena. Fortunately, some schemes have been proposed in recent years for generating such CSs. One method involves the center of mass of a laser-driven trapped ion \cite{Ref11}, while another utilizes an optical cavity \cite{Ref12}. Therefore, it is theoretically possible to control and prepare adjustable sources of SCSs, making it feasible to experimentally implement analog models based on these states.

To investigate curvature effects in quantum optics, recently some analog models introduced by employing the SCSs. For example, it has been shown that if SCSs served as one input state of a 50:50 beam splitter and vacuum as another one, we encountered with output beams with curvature-dependent quantum statistical properties \cite{Ref13}. In another analog model, the interaction of the SCSs and a three-level lambda type atom investigated. It is shown that by increasing the curvature, the collapse and revival of Rabi oscillations occurs in shorter time intervals, which confines the time dilation near massive bodies \cite{Ref12}.

In the present contribution, we adopt an analog model of the general relativity to investigate the spatial curvature effects on another important quantum optics process: the stimulated parametric down conversion (Stim.PDC). Stim.PDC is a second-order nonlinear process that has recently attracted considerable interest as a source of entangled photon pairs \cite{Ref14}. We use two-mode SCSs \cite{Ref15} as the curvature-dependent input beams for Stim.PDC and analyze the impact of spatial curvature on the quantum properties of the resulting twin photon output beams, such as entanglement, photon number statistics and etc.

To investigate entanglement between the output modes, we calculate the linear entropy as a function of spatial curvature. The results show that the amount of the entanglement can be controlled by selecting appropriate two-mode SCSs and optimizing the parameters of the Stim.PDC process. Furthermore, to explore the effects of curvature on the quantum statistical properties of Stim.PDC, we analyze the mean photon number, the Mandel parameter of the output states, and their correlation-functions.

The paper is organized as follows. In Section 2, we briefly review the SCSs and a scheme for their experimental generation. In Section 3, after a short review of the Stim.PDC process, we propose our scheme for stimulated curvature-dependent PDC. The quantum statistical properties of output states of the stimulated curvature-dependent PDC are investigated in section 4. Finally, the summary and concluding remarks are given in Section 5.

\section{Coherent states on the sphere}
As mentioned in the introduction, we use sphere coherent states (SCSs) to investigate the effects of spatial curvature on the Stim.PDC process. To set the stage for this analysis, we first provide a brief review of these two-mode states.

\subsection{Two-mode Sphere Coherent states}
The problem of the harmonic oscillator on the surface of a sphere was investigated in Ref \cite{Ref16}. In that work, the authors demonstrated that the algebra of the two- dimensional oscillator on a sphere of radius $R$ can be identified as a deformed version of $su$($2$) algebra, $su_{\lambda}$($2$), as follows:
\begin{equation}
	[\hat{J}_{0}, \hat{J}_{\pm}] = \pm \hat{J}_{\pm}, \; \; \; [\hat{J}_{+}, \hat{J}_{-}]= 2 \hat{J}_{0} h(\lambda, M, \hat{J}_{0}),
	\label{Eq:1}
\end{equation}
where,
\begin{align}
	h(\lambda, M, \hat{J}_{0}) &= 1+ \lambda \left(1+\frac{\lambda}{4} \right)^{1/2}(M+1)  \nonumber \\
	& - \lambda^2 \left[ 2 \hat{J}^{2}_{0} - N \left(\frac{M}{2} +1 \right)- \frac{1}{4}\right].
	\label{Eq:2}
\end{align}
Here, $\lambda = (1/R^{2})$ represents the curvature of the sphere, and $M$+$1$ is the dimension of the associated Hilbert space. In the flat space limit, $\lambda \rightarrow 0$, the function $h(\lambda, M, \hat{J}_{0})$ reduces to $1$, and the deformed $su_{\lambda}(2)$ algebra, reverts to the standard $su(2)$ algebra.

A generalized two-boson realization of the $su_{\lambda}(2)$ algebra is proposed in Ref.\cite{Ref14}. This  realization involves a nonlinear (f-deformed) Schwinger representation as follows:
\begin{align}
	\hat{J}_{+} =&\ \sqrt{c_{1}(\lambda) + c_{2}(\lambda) [\hat{n}^{2}_{1} + \hat{n}_{2}(\hat{n}_{2}+2)]} \: \: \hat{a}^{\dagger}_{1} \hat{a}_{2},  \nonumber \\	    	
	\hat{J}_{-} =&\ \hat{a}_{1} \hat{a}^{\dagger}_{2} \sqrt{c_{1}(\lambda) + c_{2}(\lambda) [\hat{n}^{2}_{1} + \hat{n}_{2}(\hat{n}_{2}+2)]}, \nonumber \\
	\hat{J}_{0} =&\ \frac{1}{2} (\hat{n}_{1} - \hat{n}_{2}),
	\label{Eq:3}
\end{align}
where,
\begin{equation}
	c_{1}(\lambda) = 1 + \lambda \left( 1 + \frac{\lambda}{4} \right)^{1/2} (M+1) +\lambda^{2} \left[ M \left( \frac{N}{2} + 1 \right) + \frac{1}{4}
	\right],
	\label{Eq:4}
\end{equation}
and,
\begin{equation}	
	c_{2}(\lambda) = - \frac{1}{2} \lambda^{2}.
	\label{Eq:5}
\end{equation}
Here, $\hat{a}_{j}^{\dagger}$ and $\hat{a}_{j}$ are the creation and annihilation operators corresponding to the $j$th mode, respectively, and $\hat{n}_{j} = \hat{a}_{j}^{\dagger} \hat{a}_{j}$. It is clear that in the flat limit, $c_{1}(\lambda) \rightarrow 1$ and $c_{2}(\lambda) \rightarrow 0$, and thus, the above nonlinear two-boson realization of $su_{2}(\lambda)$ reduces to the standard Schwinger realization of $su(2)$ \cite{Ref16}. The two-mode SCSs for this nonlinear bosonic realization are constructed as \cite{Ref14}:
\begin{equation}
	|z;\lambda,M \rangle = \mathcal{N}^{-1\slash 2} \sum_{m=0}^{M} \sqrt{\binom{M}{m}} \; 	[g(\lambda,m)]! \;z^{m} |m, M-m\rangle,
	\label{Eq:6}
\end{equation}
where $\left| n,m \right\rangle =\left| n \right\rangle \otimes \left| m \right\rangle$ $\in$ $\hat{\mathcal{H}_{1}} \otimes \hat{\mathcal{H}_{2}}$, and $z$ is a complex number. The normalization factor $\mathcal{N}$ is given by:
\begin{equation}
	\mathcal{N} = \sum_{m=0}^{M} \binom{M}{m} ([g(\lambda,m)]!)^{2} |z|^{2m},
	\label{Eq:7}
\end{equation}
by definition,
\begin{equation}
	[g(\lambda , 0)]! = 1, \: \: \: \: \: [g(\lambda, m)]! = g(\lambda, m) [g(\lambda, m-1)]!.
	\label{Eq:8}
\end{equation}
The curvature-dependent function $g(\lambda, m)$ defined as:
\begin{equation}
	g(\lambda,m) = \sqrt{(\lambda (M+1-m))+ \sqrt{1 + \frac{\lambda^{2}}{4}}} \sqrt{(\lambda m + \sqrt{1+ \frac{\lambda^{2}}{4}})}.
	\label{Eq:9}
\end{equation}
In the flat limit ($\lambda \rightarrow 0$), $g(\lambda, m) \rightarrow 1$ and the above two-mode SCs reduce to the two-mode CSs for bosonic realization of the $su(2)$ algebra \cite{Ref18}.

\subsection{\label{Sec:2.2}Experimental realization of two-mode SCSs}
In Ref.\cite{Ref19}, a scheme was proposed to generate a class of two-mode field states in a cavity. A Raman-coupled three-level atomic system interacts with a two-mode cavity field. By employing appropriate  off-resonance condition in the cavity, the anti-Stokes modes are eliminated, ensuring that the total number of photons in both modes remains constant. The atomic system consists of a series of three-level $\Lambda$-type atoms prepared in a superposition of their two non-degenerate ground states, expressed as $\left| 3 \right\rangle  + \epsilon_{k} \left| 1 \right\rangle$. Here, $\left| 1 \right\rangle $ and $ \left| 3 \right\rangle$ are two non- degenerate ground states of the atoms,  $\epsilon_{k}$ is the complex amplitude coefficient for the $k$th atom and $E_{3}<E_{1}$. These atoms interact with a two-mode cavity field,  where the first mode is initially in a vacuum state, and the second mode contains $M$ photons. Each injected atom increases the photon number in the first mode by one while simultaneously destroying a photon from the second mode. It is assumed that after the $(k-1)$th atom has passed and just before the injection of the $k$th atom, the cavity field is in the state $ \left| \phi^{(k-1)} \right\rangle  = \sum_{m=0}^{M} \phi^{(k-1)}_{m} \left| m, M-m \right\rangle $. As soon as an atom exits the cavity, it is detected whether the atom is in the state $\left| 3 \right\rangle$ or $\left| 1 \right\rangle$. If the atom is detected in the ground state, the process should continue to allow for additional energy transfer from the atom to the field in order to achieve the desired state. Conversely, if the atom is found in the excited state, the process must be repeated. The new coefficients $\phi^{(k)}_{m}$ for the field state after the exit of $k$th atom, represented as $\left| \phi^{(k)} \right> = \sum_{m=0}^{M} \phi^{(k)}_{m} \left| m, M-m \right>$, are given in terms of the old coefficients  $\phi^{(k-1)}_{m}$ according to a recurrence relation \cite{Ref19}.

Now, by using a similar approach, one can prepare the two-mode SCSs, Eq. (\ref{Eq:6}). To achieve this, it is essential to determine a combination of $M$ number states represented as $ \left| \phi^{(k-1)} \right\rangle  = \sum_{m=0}^{M} \phi^{(k-1)}_{m} \left| m, M-m \right\rangle $. This combination should lead to the state $ \left| z; \lambda , M \right\rangle$ after the $M$th atom, which is prepared in an appropriate internal state $ \left| 3 \right\rangle + \epsilon_{M} \left| 1 \right\rangle$, passes  through the cavity and is detected in the ground state. In accordance with the approach described in Ref.\cite{Ref20}, we can derive a characteristic polynomial equation of order $M$ for $\epsilon_{M}(\lambda)$. After solving this equation, we identify the smallest root among $M$ solutions and select it as the value of  $\epsilon_{M}(\lambda)$. Once we have determined $\epsilon_{M}(\lambda)$, we can calculate the corresponding set of $\phi^{(M-1)}_{m}$ coefficients. In the subsequent step, we designate $\left| \phi^{(M-1)}\right\rangle $ as the new target state, which we aim to achieve by sending $M-1$ atoms through the cavity. For the state $ \left| \phi^{(M-1)} \right\rangle  $ we can perform similar calculations as we did for the state $\left| z; \lambda , M \right\rangle$. This will allow us to determine the parameter  $\epsilon_{M-1}(\lambda)$ and to express the state $ \left| \phi^{(M-1)} \right\rangle$ in terms of the $M-1$ coefficients $ \phi^{(M-2)}_{m}$. The calculations are repeated until we arrive at the initial field state. The sequence of $\lambda$-dependent complex numbers $\epsilon_{1}(\lambda),\epsilon_{2}(\lambda), ...,\epsilon_{M}(\lambda)$ specifies the internal states for a series of $M$ atoms that need to be injected into the cavity. This process enables the generation of the desired two-mode SCSs in a two-mode resonator \cite{Ref12}.

\section{Stimulated curvature-dependent PDC}
In this section, we intend to employ two-mode SCSs as the input beams for the Stim.PDC process, thereby introducing curvature dependence. To proceed, we will first briefly review the Stim.PDC process.

\subsection{Stimulated parametric down-conversion}
PDC is an optical process that has revolutionized the field of quantum optics \cite{Ref21, Ref22}. Also, the source of this process provides one of the most important experimental tools for investigating quantum entanglement in the quantum information and quantum computation \cite{Ref23}. This process involves the coupling between three optical modes via a second order nonlinearity inside a nonlinear crystal \cite{Ref24}. In this nearly elastic process, a single pump photon is transformed into two lower-energy photons, known as the signal and idler, while conserving both energy and momentum. In this process, the generation of entangled signal and idler photons occurs spontaneously, a phenomenon that cannot be explained by classical nonlinear optics. Additionally, observing a high coincidence count between signal and idler photons at separate detectors indicates the violation of classical inequalities, which all classical states adhere to \cite{Ref25}. Additional quantum signatures arise from the strong correlations between the transverse components of the wave vectors of the signal and idler photons \cite{Ref21}. Beside, the intrinsic quantum properties exhibited by PDC sources in their unmodified form, entanglement can be prepared and detected in additional degrees of freedom, most notably polarization. Various techniques exist to generate polarization-entangled photon pairs \cite{Ref26, Ref27} which serve as one of the fundamental experimental resources in quantum information science.

On the other hand, the PDC process can be stimulated by  inserting at least one extra photon beam, where the frequency of this beam matches either the signal or the idler beam. \cite{Ref28}. This  process, known as Stim.PDC, produces light beams that are correlated in both intensity and phase \cite{Ref29, Ref30}, making it an essential tool in quantum information science \cite{Ref31}.

Assuming that the pump intensity is high enough, we can consider the corresponding field intensity classically and write the Hamiltonian of PDC as follows \cite{Ref32}:
\begin{equation}
	H_{pdc}= i\hbar \eta A_{p} \hat a_{s}^{\dagger} \hat a_{i}^{\dagger} - i\hbar \eta^{*} A_{p}^{*} \hat a_{s} \hat a_{i},
	\label{Eq:10}
\end{equation}
where $a_{s}$ ($a_{s}^{\dagger}$) and $a_{i}$ ($a_{i}^{\dagger}$) are the annihilation (creation) operators for the signal and idler modes, respectively. Here, $\eta$ characterizes the non-linear interaction, while $A_{p}$ represents the amplitude of the classical strong pump field. Consequently, the unitary time evolution operator associated with this Hamiltonian can be expressed as follows:
\begin{equation}
	\hat{U}_{pdc}(t) \equiv e^{-i \hat{H}_{pdc}t/\hbar}=e^{\left(\tau \hat a_{s}^{\dagger} \hat 	a_{i}^{\dagger}-\tau^{*} \hat a_{s} \hat a_{i} \right) },
	\label{Eq:11}
\end{equation}
where $\tau = \eta A_{p} t$. Following \cite{Ref32}, we can write this operator as follows:
\begin{equation}
	\hat{U}_{pdc}(t) = e^{\tanh{(r)} e^{i \theta}  \hat L_{+}} \: e^{-2\ln{\cosh{(r)}} \hat L_{0} } \: e^{-\tanh{(r)} e^{- i \theta} \hat L_{-}},
	\label{Eq:12}
\end{equation}
where we have defined $\hat L_{+} = \hat a_{s}^{\dagger} \hat a_{i}^{\dagger}$, $\hat L_{-} = \hat a_{s} \hat a_{i}$ and $\hat L_{0} = 1/2 (\hat a_{s}^{\dagger} \hat a_{s} +\hat a_{i}^{\dagger} \hat a_{i} + 1)$. Also, the complex coefficient $\tau$ is written as $r e^{i \theta}$. It can be easily shown that these new operators construct the following $su(1,1)$ algebra:
\begin{equation}
	\left[\hat L_{+}, \hat L_{-} \right] = -2 \hat{L}_{0}, \: \: \left[\hat L_{0}, \hat L_{\pm} \right] = \hat{L}_{\pm}.
	\label{Eq:13}
\end{equation}

\subsection{Stimulated curvature-dependent PDC}
\begin{figure}[h!]
	\begin{center}
		\includegraphics[width=10 cm]{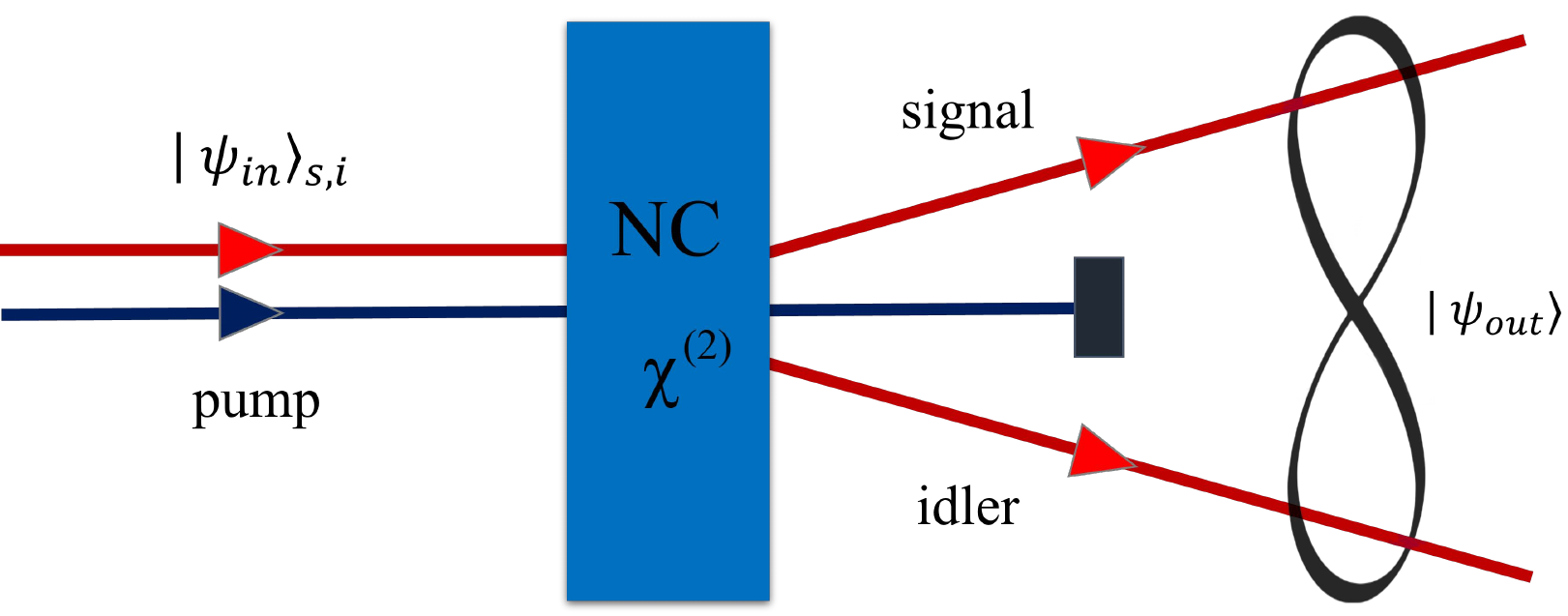}
		\caption{\small Stim.PDC process, as a seeded PDC by using initial  $|\psi_{in} \rangle_{s,i}$ state. The output entangled output state is shown as $|\psi_{out} \rangle$}
		\label{Fig:0}
	\end{center}
\end{figure}

In the spontaneous PDC process \cite{Ref24}, the initial states of both the signal and idler modes are typically considered to be in the vacuum state. However, in the Stim. PDC process, the initial states of the signal and idler modes can be seeded, as shown in Fig. \ref{Fig:0}. In this section, we assume that the initial states of the signal and idler modes are prepared in SCSs, which their generations are explained in section \ref{Sec:2.2}. By seeding the Stim.PDC process by a two-modes SCS as the initial state, the output state of the Stim.PDC process will be given by:
\begin{eqnarray}
	|\psi_{out}\rangle &=& \hat{U}_{pdc}(t) \:|\psi_{in}\rangle_{s, i} \nonumber \\
	&=& \mathcal{N}^{-1/2}  \sum_{m=0}^{M} \sqrt{\binom{M}{m}} \: [g(\lambda, m)]! \: z^{m} \: \hat{U}_{pdc}(t) \: |m \rangle_{s} |M-m \rangle_{i}
	\label{Eq:14}
\end{eqnarray}
By using (\ref{Eq:8}), we can obtain the  output states as the following:
\begin{align}
	|\psi_{out}\rangle &= \mathcal{N}^{-1/2} \sum_{m=0}^{M} \sum_{q=0}^{\beta} 	\sum_{p=0}^{\infty} \: (-1)^{q} C_m \: z^{m} \:  \sqrt{\binom{m}{q}\binom{M-m}{q} \binom{m-q+p}{p}} \nonumber \\
	&\times \sqrt{ \binom{M-m-q+p}{p}} \: e^{i(p-q)\theta} \: [\tanh(r)]^{p+q} \: [\cosh(r)]^{-(M-2q+1)} \nonumber \\
	&\times |m-q+p\rangle_{s} |M-m-q+p \rangle_{i} \: ,
	\label{Eq:15}
\end{align}
with:
\begin{equation}
	C_{m}=\sqrt{\binom{M}{m}} \: [g(\lambda, m)]!,
	\label{Eq:16}
\end{equation}
and:
\begin{equation}
	\beta = Min\{M-m, m\}.
	\label{Eq:17}
\end{equation}

From Eq. (\ref{Eq:12}), it is evident that the effect of non-linear crystal on the seeded SCSs is managed by the pump power $A_{p}$, the interaction strength within the crystal $\eta$, and the interaction time $t$. Moreover, it is clear from Eq. (\ref{Eq:15}) that adjusting the value of $r$ (which can be controlled through the crystal length and the pump power) effectively limits the number of efficient creation and annihilation operations. For a specific crystal with fixed length, the parameter $r$ is a function of the pump power. Hence we call it the $pump \: power \: parameter$ through the paper.

\section{Quantum Statistical Properties of the Output States}
In this section, we shall proceed to explore the influence of curvature on some quantum-information and quantum-statistical properties of the output states. Specifically, we study some of the key aspects, including the entanglement, the mean photon number, the Mandel parameter and the cross-correlation function.

\subsection{Von-Neumann linear entropy}
In the following, we construct the density matrix and calculate the Von-Neumann linear entropy \cite{Ref33}, as a measure of entanglement, to examine the effects of spatial curvature on the entanglement between our output states. The linear entropy for our bipartite output state $\rho_{s i}$ is defined as \cite{Ref34}:
\begin{equation}
	S(\rho_{s}) = 1-Tr_{s}(\rho_{s}^{2}),
	\label{Eq:18}
\end{equation}
where the reduced density operator $\rho_{s}$ is given by:
\begin{equation}
	\rho_{s} = Tr_{i}[|\psi_{out}\rangle \langle \psi_{out}|].
	\label{Eq:19}
\end{equation}

\begin{figure}[h!]
	\begin{center}
		\includegraphics[width=10 cm]{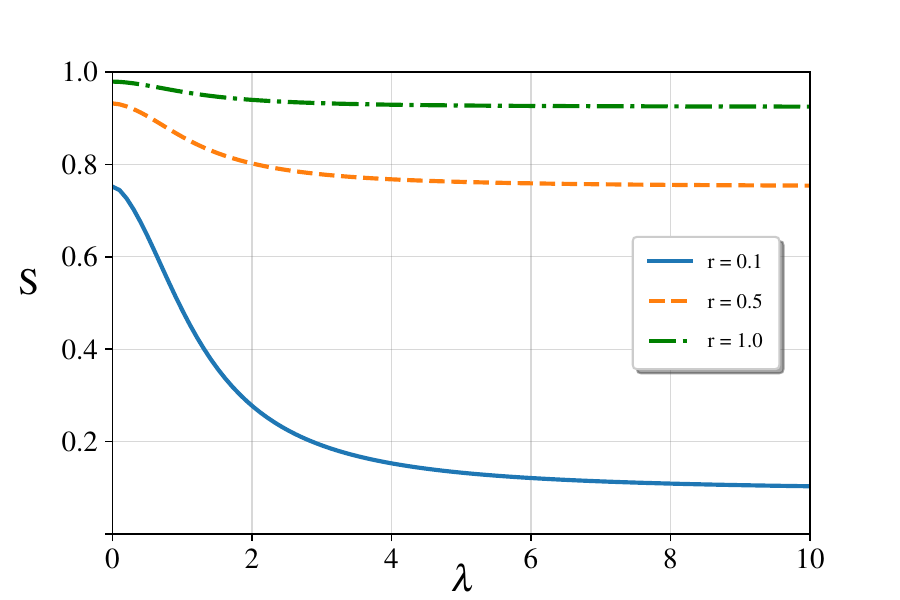}
		\caption{\small Von-Neumann linear entropy as a function of space curvature $\lambda$ for $M=4$ and $z=1$. The solid (blue) line corresponds to $r=0.1$ , the dashed (orange) to $r=0.5$, and the green (dash-dotted) to $r=1.0$. }
		\label{Fig:1}
	\end{center}
\end{figure}

By substituting Eq. (\ref{Eq:15}) into Eq. (\ref{Eq:19}), we can derive the linear entropy of the output states of our curvature-dependent Stim.PDC. Fig. \ref{Fig:1}, illustrates the variation of the linear entropy $S$ with respect to spatial curvature $\lambda$ with $M=4$ and $z=1$, for three different values of the pump power parameter: $r=0.1$, $0.5$ and $1$. As is seen, for $r = 0.1$, the entanglement is decreased by increasing $\lambda$. This reduction persists until $\lambda = 8$, beyond which the entanglement remains constant. However, as $r$ increases, the entanglement between two modes is significantly enhanced. Moreover, the decline due to curvature becomes less significant, and the $\lambda$ threshold at which the entanglement stabilizes shifts to lower values.

\begin{figure}[h!]
	\begin{center}
		\includegraphics[width=10 cm]{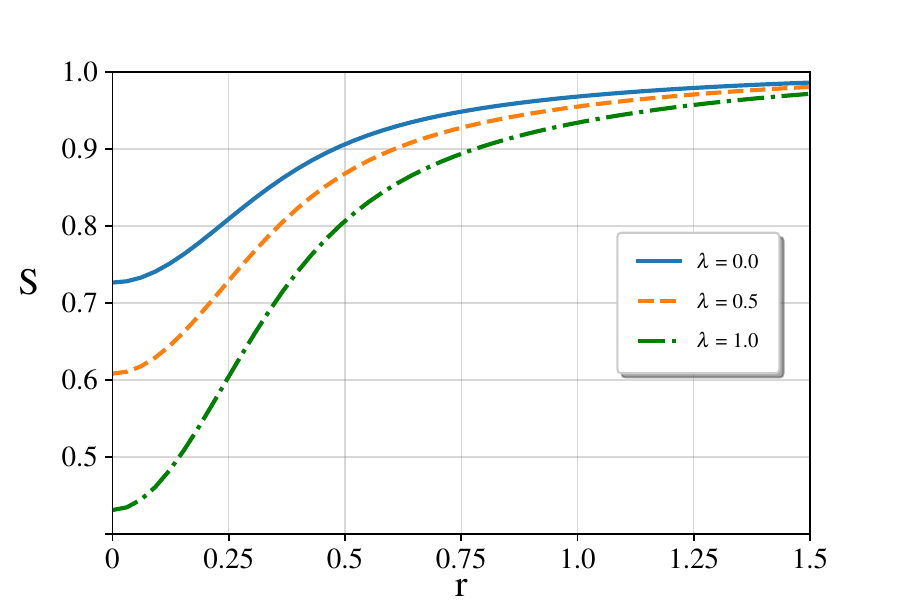}
		\caption{\small Linear entropy for SCSs versus $r$, for $M=4$ and $z=1$. The solid (blue) line corresponds to $\lambda=0$, the dotted (orange) to $\lambda = 0.5$ and the dash-dotted (green) to $\lambda = 1$.}
		\label{Fig:2}
	\end{center}
\end{figure}

In Fig. \ref{Fig:2}, we have plotted the linear entropy as a function of $r$ with $M=4$ and $z=1$ for three different spatial curvatures:  $\lambda=0, \: 0.5, \: 1$. We see that for a fixed value of curvature, increasing $r$ leads to a gradual enhancement of the entanglement, asymptotically approaching its maximum value. It is worth noting that, although at the higher curvature values the initial entanglement is smaller, by increasing $r$, the entanglement eventually converges to the same limiting value, regardless of the given curvature. Consequently, for a fixed $\lambda$, the entanglement can be controlled by adjusting the pump power parameter, the interaction efficiency, and the interaction duration time.

\begin{figure}[h]
	\begin{center}
		\includegraphics[width=10 cm]{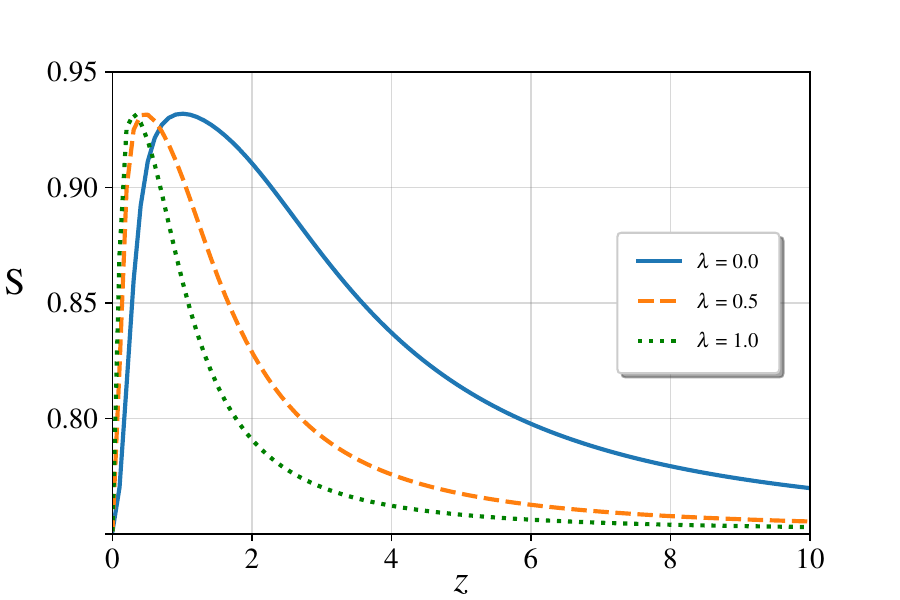}
		\caption{Linear entropy f versus $z$, for $M=4$ and $r=0.5$, with $\lambda = 0$ for solid (blue) line, $\lambda=0.5$ for dashed (orange) and $\lambda=1$ for dotted (green). }
		\label{Fig:3}
	\end{center}
\end{figure}

Another plot that can be useful for us, is the linear entropy with respect to the parameter $z$, as is shown in Fig. \ref{Fig:3}.
From Eq. (\ref{Eq:15}), it is evident that, we can manipulate the behavior of each mode by using the $z$ parameter. The results of Fig. \ref{Fig:3} reveals that, for a fixed value of $\lambda$, increasing the $z$ results entanglement enhancement up to a specific maximum value. For instance, in the flat limit, $\lambda \rightarrow 0$, this peak occurs around $z = 1$. Also, by increasing  the curvature, this critical threshold, shifts to lower $z$ values, and the reduction in entanglement occurs more rapidly.

\begin{figure}[h]
	\begin{center}
		\includegraphics[width=10 cm]{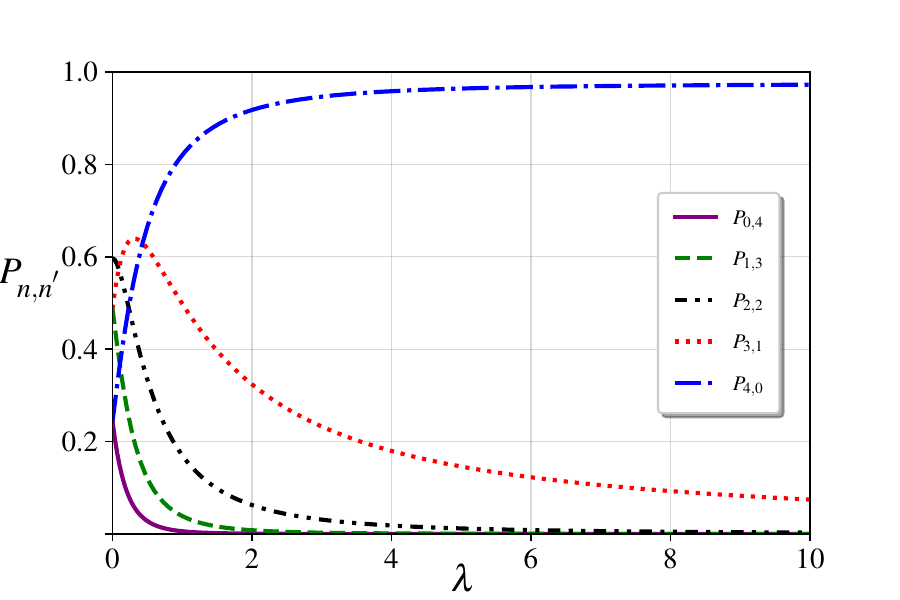}
		\caption{\small  $P_{n, n'}\equiv|R_{m-p+q,\: M-m-q+p}|^{2}$, given by Eq. (\ref{Eq:25}), as a function of $\lambda$ with $M=4$, $z=1$ and $r=0.1$.}
		\label{Fig:4}
	\end{center}
\end{figure}

For further illustration, we can rewrite Eq.(\ref{Eq:19}) as follow:
\begin{equation}
	|\psi_{out}\rangle = \sum_{m=0}^{M}\sum_{p=0}^{\beta + q} \sum_{q=0}^{\beta} R_{m-p+q,\: M-m-q+p} |m-q+p\rangle_{s} \: |M-m-q+p\rangle_{i},
	\label{Eq:20}
\end{equation}
where,
\begin{align}
	&R_{m-p+q,\: M-m-q+p} = \mathcal{N}^{\ -1/2} \: C_m \: z^{m} \: \sqrt{\binom{m}{q} \binom{M-m}{q} \binom{m-q+p}{p}} \nonumber \\
	& \: \: \times \sqrt{\binom{M-m-q+p}{p}} \: e^{i(p-q)\theta} \left( \tanh(r)\right) ^{p+q} \left(\cosh(r)\right) ^{-(M-2q+1)} ,
	\label{Eq:25}
\end{align}
is the probability amplitude that corresponds to obtaining the two-mode state $|m-q+p\rangle_{s} |M-m-q+p\rangle_{i}$ in the output state . Now, we can plot $P_{n, n'}\equiv|R_{m-p+q,\: M-m-q+p}|^{2}$ as a function of $\lambda$ to see how this probabilities change versus the curvature of space. Fig. \ref{Fig:4} shows $P_{n, n'}$ as a function of $\lambda$ with $M=4$, $z=1$ and $r=0.1$.
In this figure, we have focused exclusively on the probabilities with substantial values. It is evident that as the curvature of space increases, all probabilities approach zero, except for the $P_{4, 0}$, which corresponds to the signal mode being fully populated, while simultaneously the idler mode is completely depleted. This observation suggests that increasing curvature enhances the distinguishability between the modes. Consequently, in agreement with Fig. \ref{Fig:1}, it is expected that increasing $\lambda$ will reduce the degree of entanglement between the two-modes.

\subsection{Photon-number distribution}
\begin{figure}[ht]
	\centering
	
	% First row
	\begin{subfigure}{0.49\textwidth}
		\centering
		\includegraphics[width=\textwidth]{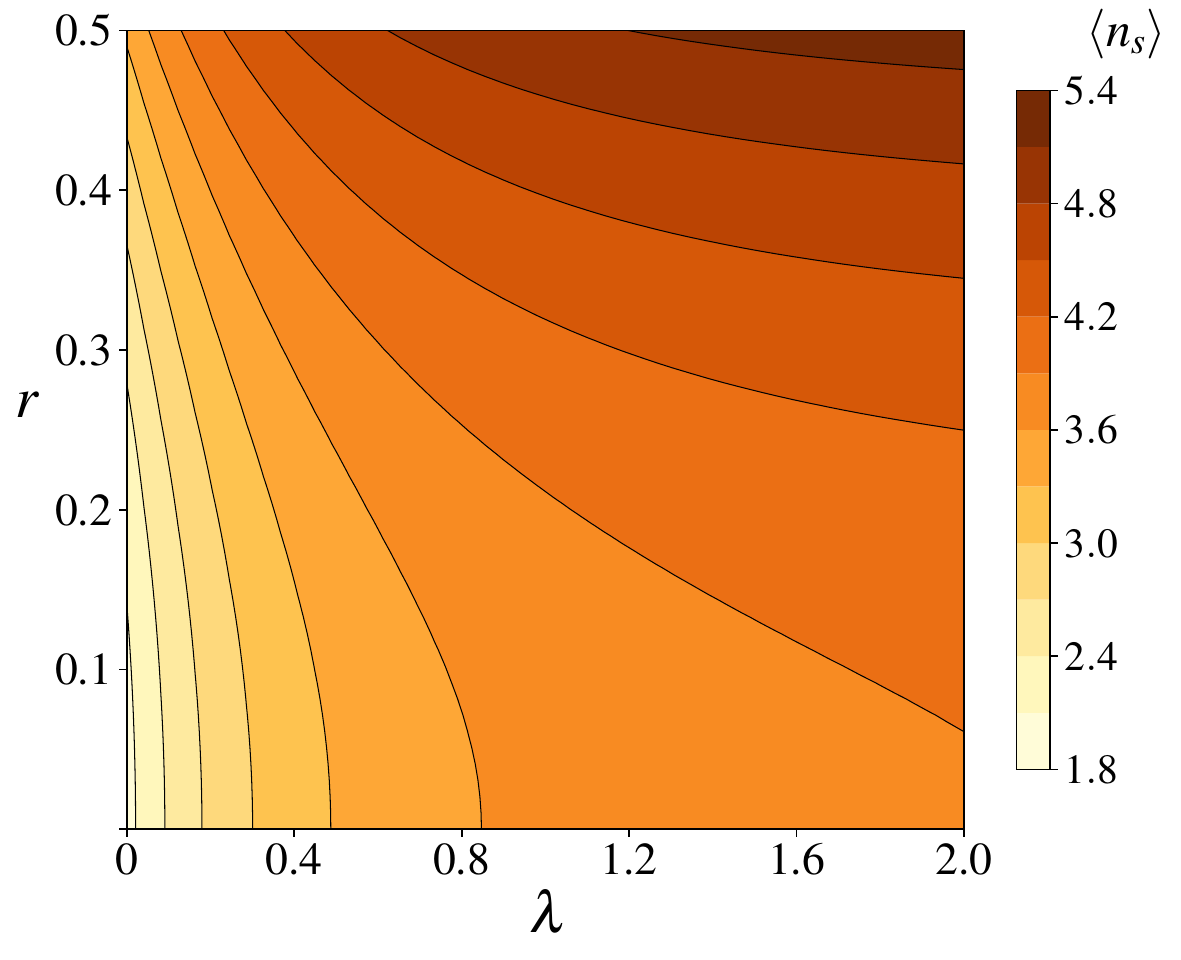}
		\caption{}
		\label{fig:sub5.1}
	\end{subfigure}
	\hfill
	\begin{subfigure}{0.49\textwidth}
		\centering
		\includegraphics[width=\textwidth]{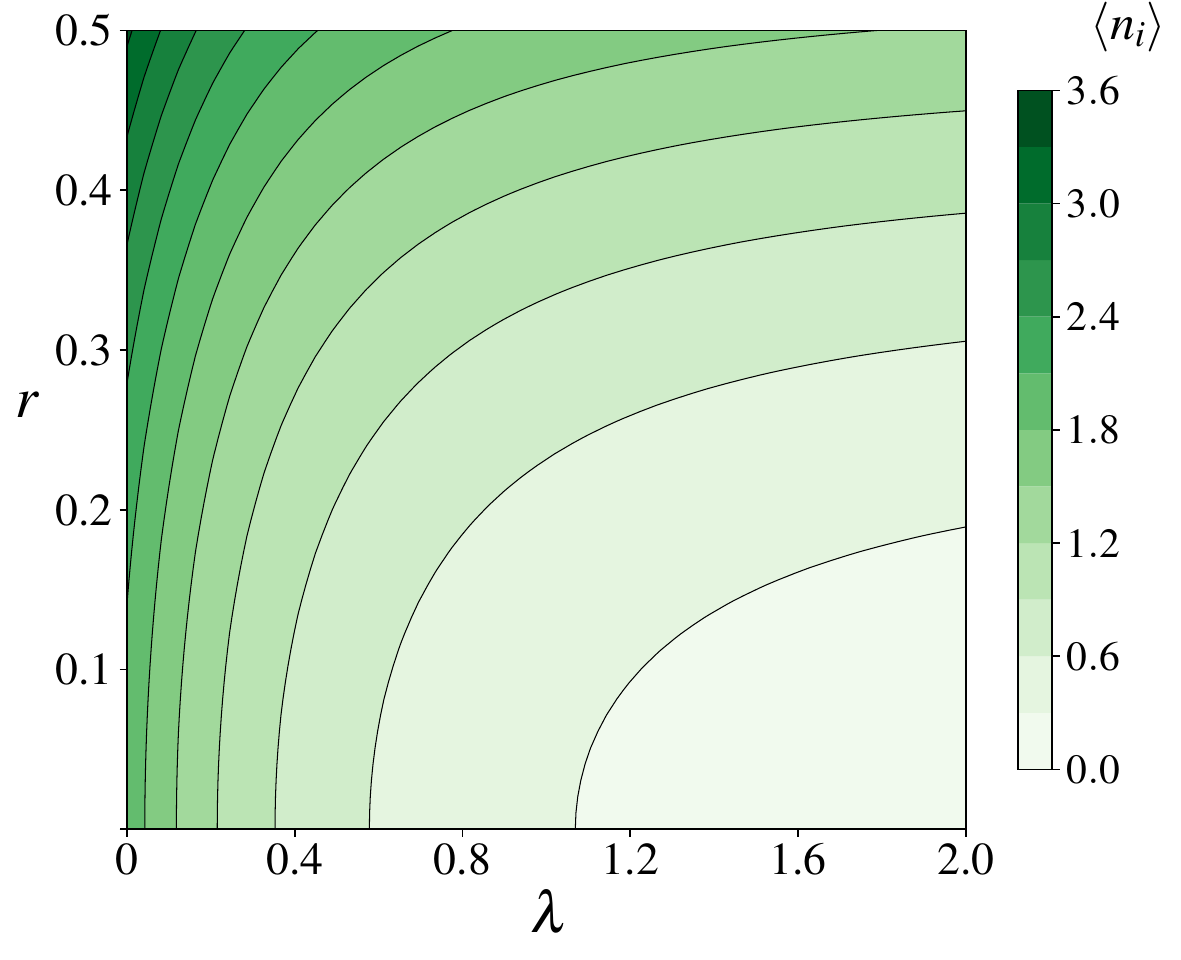}
		\caption{}
		\label{fig:sub5.2}
	\end{subfigure}
	
	\vspace{0.1cm}  % Optional space between rows
	
	% Second row
	\begin{subfigure}{0.49\textwidth}
		\centering
		\includegraphics[width=\textwidth]{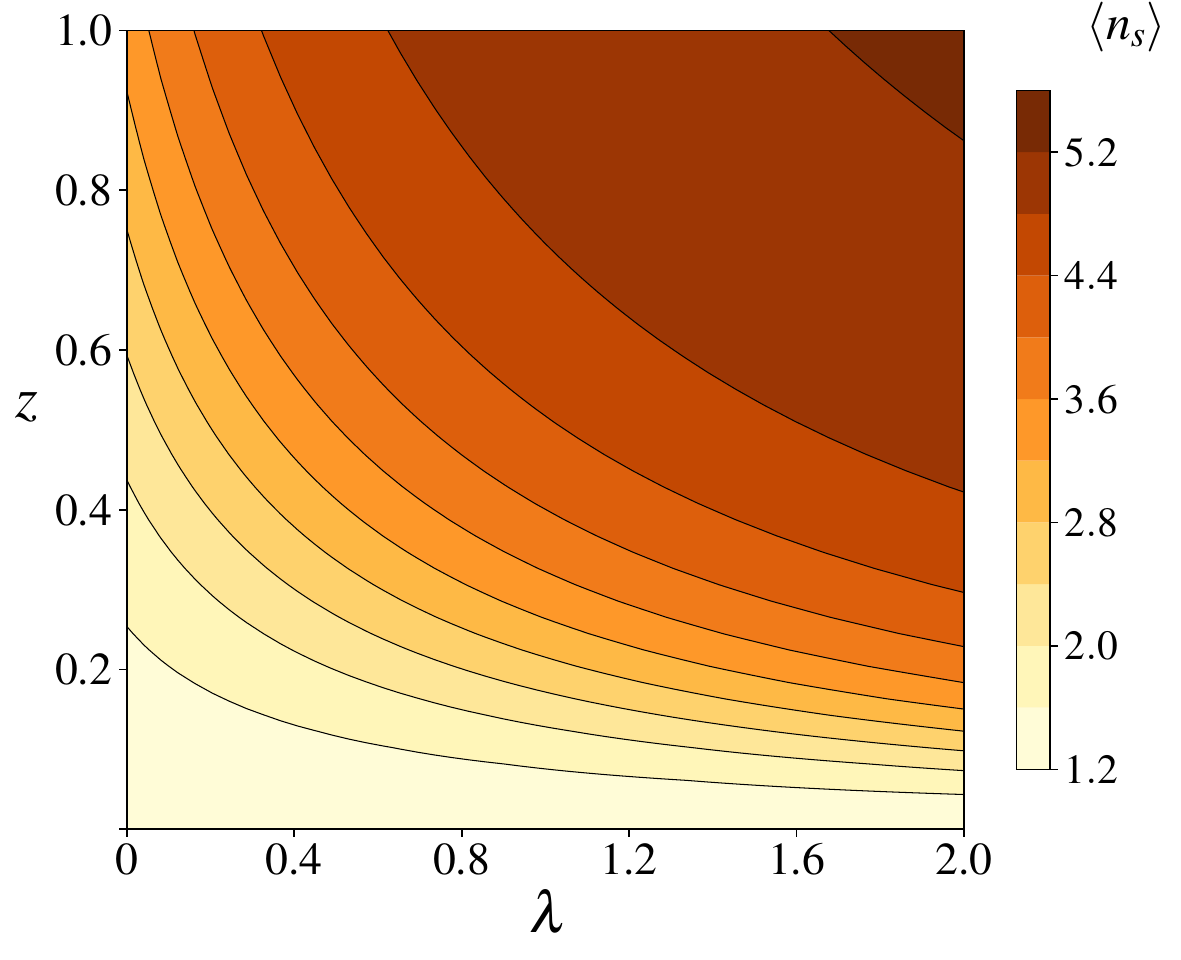}
		\caption{}
		\label{fig:sub5.3}
	\end{subfigure}
	\hfill
	\begin{subfigure}{0.49\textwidth}
		\centering
		\includegraphics[width=\textwidth]{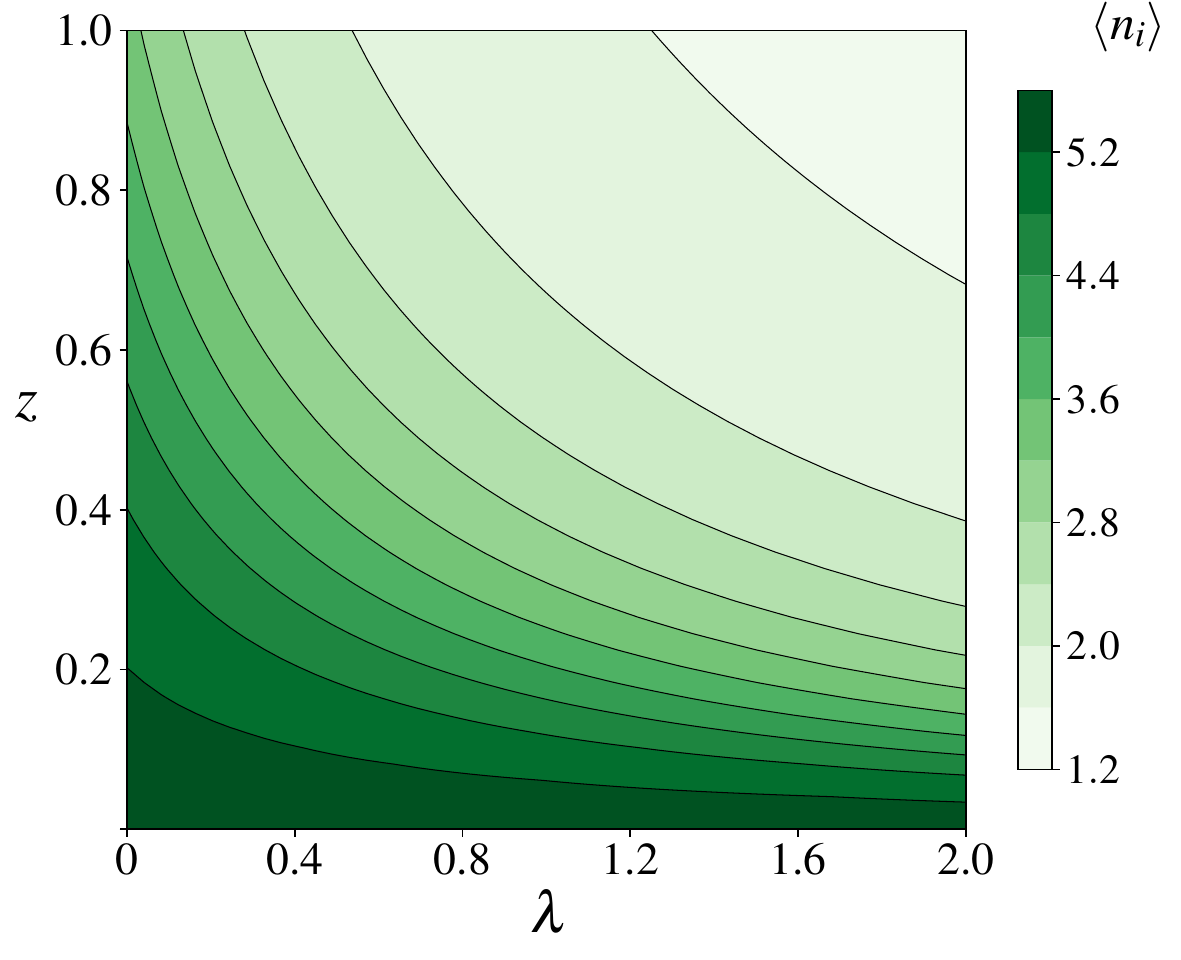}
		\caption{}
		\label{fig:sub5.4}
	\end{subfigure}
	
	\caption{Contour plots of the mean photon number for the signal (left column) and idler (right column) modes with $M = 4$. Contours (a) and (b) correspond to a fixed $z = 1$, while (c) and (d) depict the case with a fixed $r = 0.5$.}
	\label{Fig:5}
\end{figure}

To provide new insights of the problem, we investigate the behavior of the mean number of photons in each mode of the output signal and idler states. In Figs. \ref{fig:sub5.1} and \ref{fig:sub5.2} we have plotted the contours of these numbers as a function of $\lambda$ and $r$ with $z=1$, and in Figs. \ref{fig:sub5.3} and \ref{fig:sub5.4} as a function of $\lambda$ and $z$ with $r = 0.5$. According to these plots, for a fixed value of $r$ and choosing $z=1$, as the curvature increases, photons tend to occupy the signal mode and leave the idler one. This behavior is in good agreement with the results presented in Fig.(\ref{Fig:4}). Also by increasing $r$, the population of each mode will be increase, but, the whole behavior is the same. It is obvious that, for flat space, the population of two modes are equal, and we cannot distinguish two modes from each other. On the other side, for $r = 0.5$, increment in $z$ results  in increment for signal population and decrement in idler one.

\begin{figure}[ht]
	\centering
	
	% First row
	\begin{subfigure}{0.49\textwidth}
		\centering
		\includegraphics[width=\textwidth]{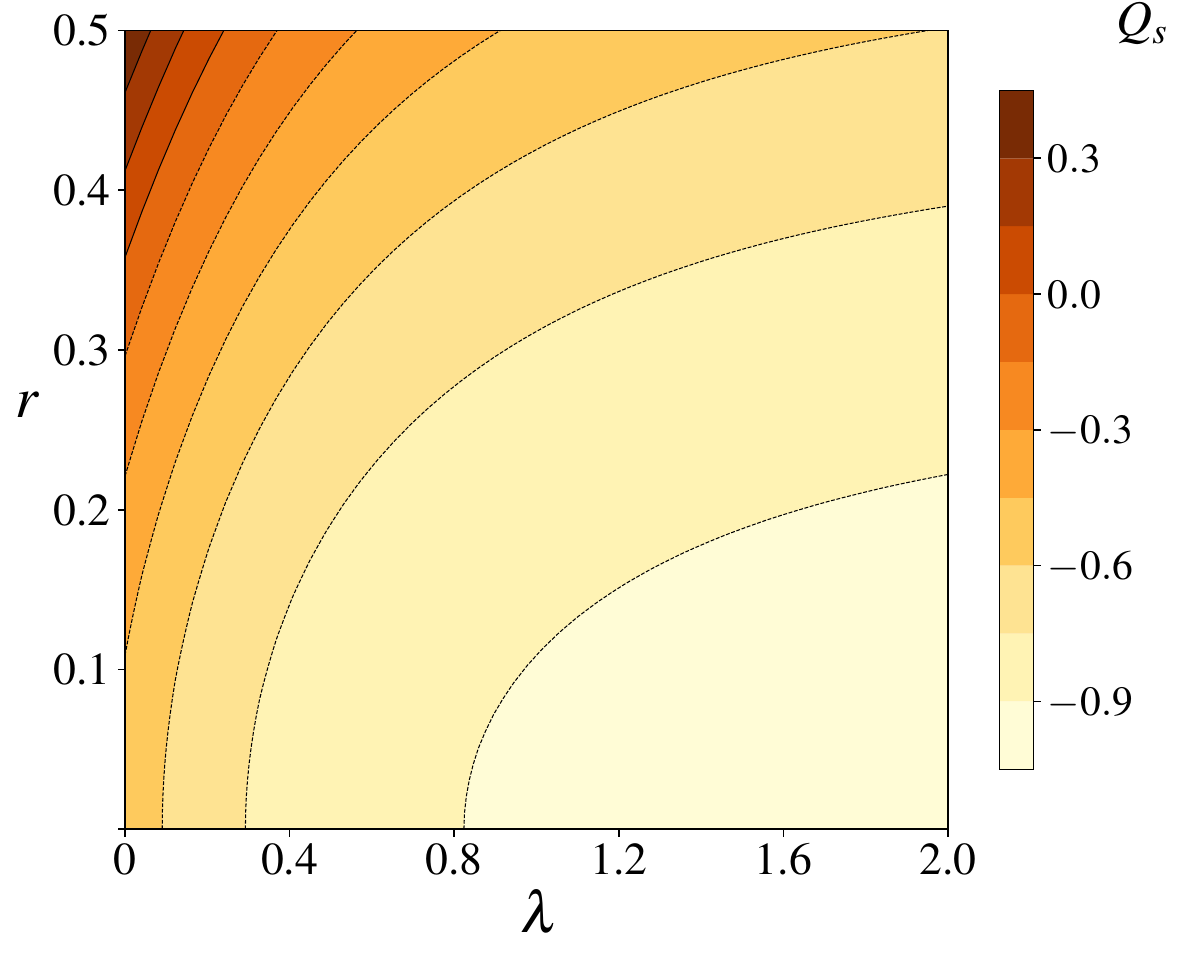}
		\caption{}
		\label{fig:sub6.1}
	\end{subfigure}
	\hfill
	\begin{subfigure}{0.49\textwidth}
		\centering
		\includegraphics[width=\textwidth]{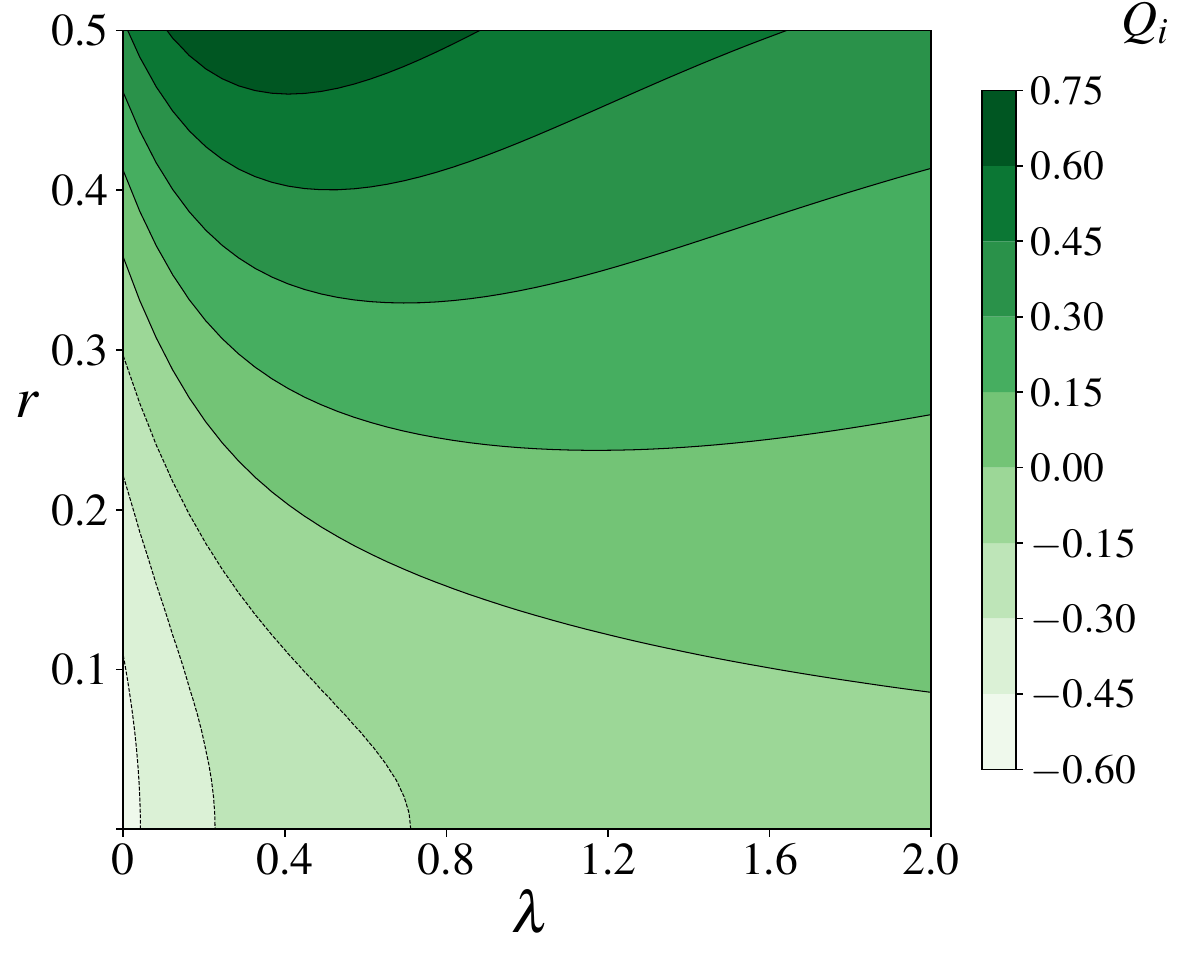}
		\caption{}
		\label{fig:sub6.2}
	\end{subfigure}
	
	\vspace{0.1cm}  % Optional space between rows
	
	% Second row
	\begin{subfigure}{0.49\textwidth}
		\centering
		\includegraphics[width=\textwidth]{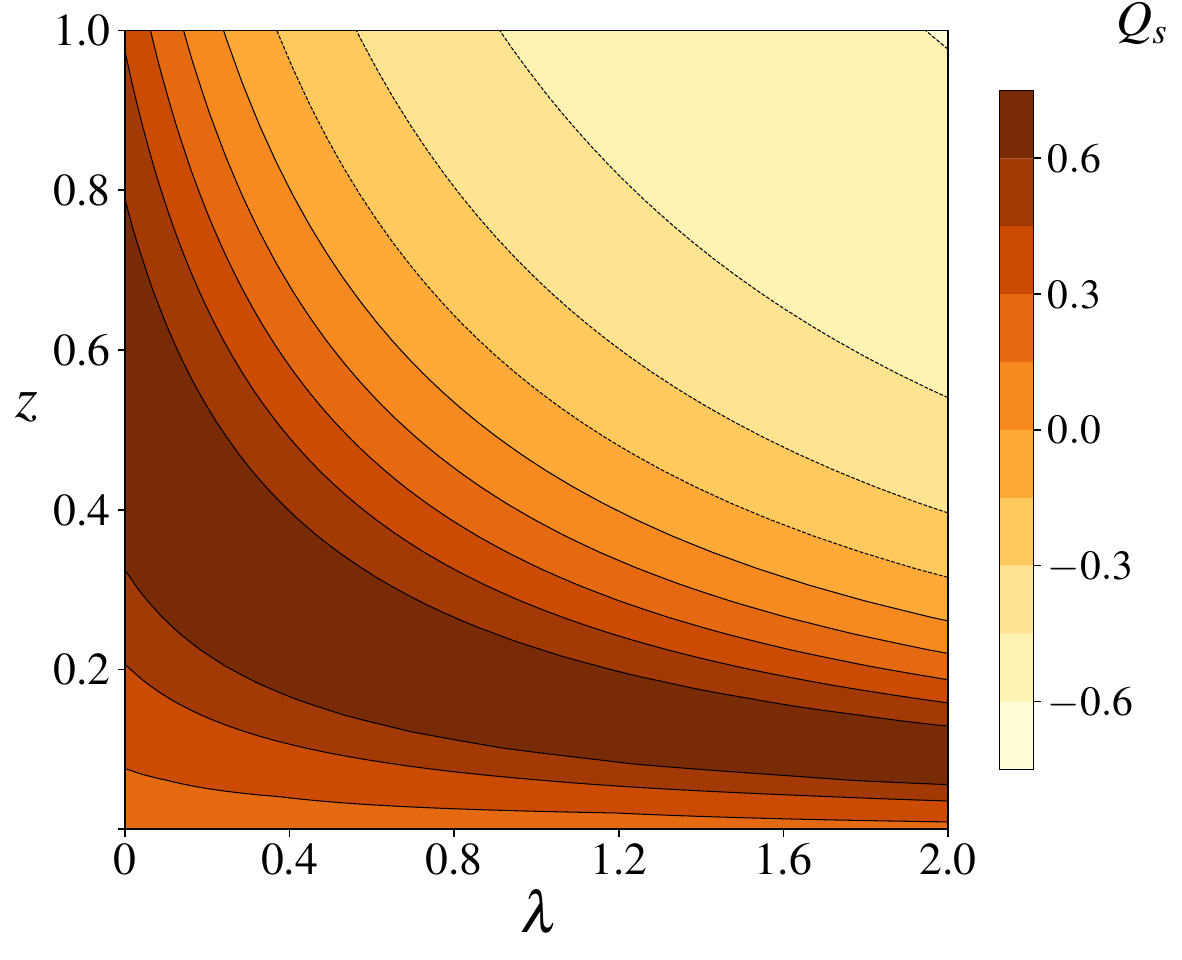}
		\caption{}
		\label{fig:sub6.3}
	\end{subfigure}
	\hfill
	\begin{subfigure}{0.49\textwidth}
		\centering
		\includegraphics[width=\textwidth]{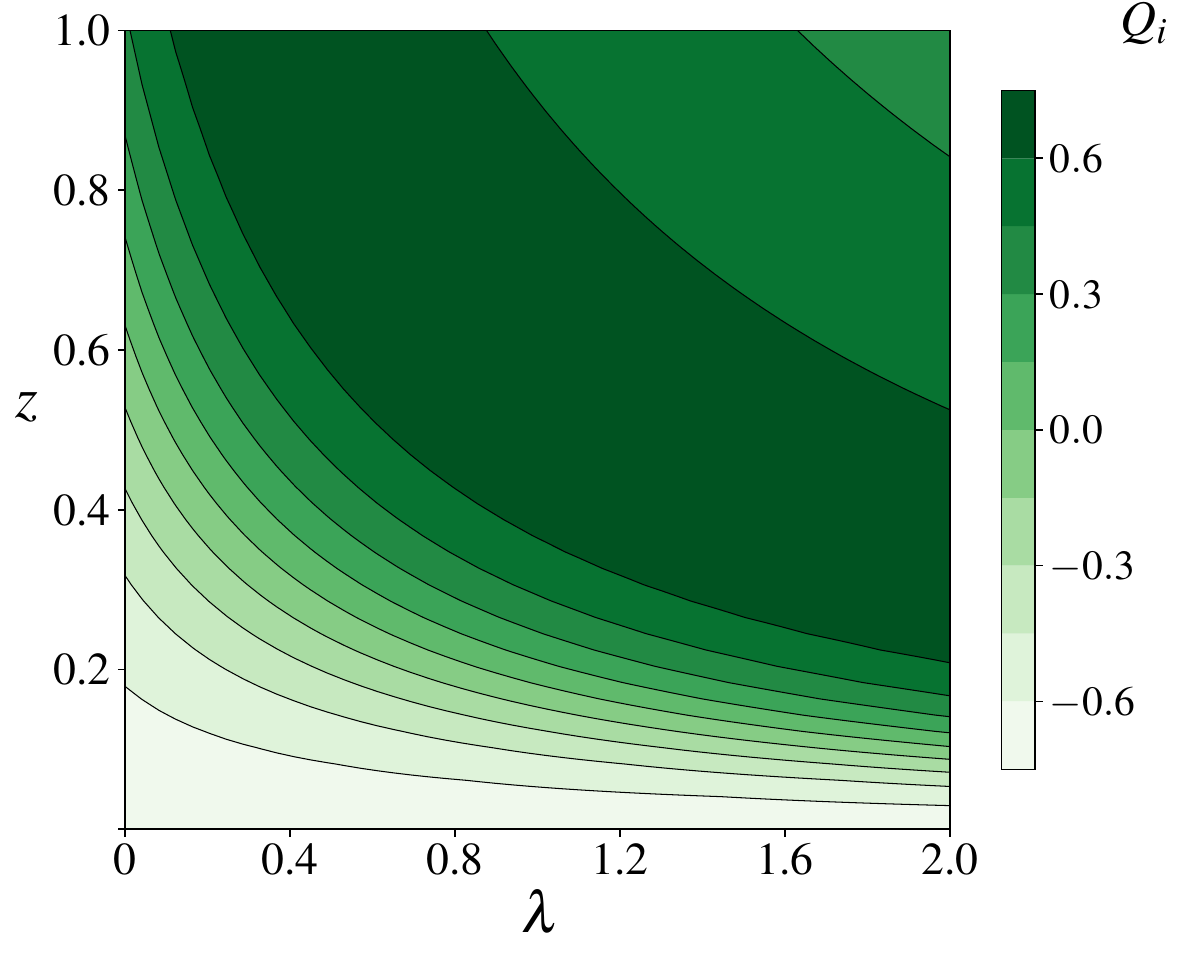}
		\caption{}
		\label{fig:sub6.4}
	\end{subfigure}
	
	\caption{Contour plots of the Mandel parameter for the signal (left column) and idler (right column) modes with $M = 4$. Contours (a) and (b) correspond to a fixed $z = 1$, while (c) and (d) depict the case with a fixed $r = 0.5$.}
	\label{Fig:6}
\end{figure}

Next parameter that can help us to understand the role of space curvature is the Mandel parameter \cite{Ref34}:
\begin{equation}
	Q_{o} = \frac{\Delta \hat{n}_{o} - \langle \hat{n}_{o}\rangle}{\langle \hat{n}_{o} \rangle},
	\label{Eq:22}
\end{equation}
with $o= s, i$. This parameter reveals a Poissonian behavior for output state if $Q_{o}=0$, a sub-Poissonian (photon bunching) one if $Q_{o}<0$, and a super-Poissonian (photon anti-bunching) if $Q_{o}>0$ \cite{Ref34}. In Figs. \ref{fig:sub6.1} and \ref{fig:sub6.2} we have plotted the contours of the Mandel parameter as a function of $\lambda$ and $r$ with $z=1$, and in Figs. \ref{fig:sub6.3} and \ref{fig:sub6.4} as a function of $\lambda$ and $z$ with $r = 0.5$. The plots clearly show that, for a fixed value of $r$ and choosing $z = 1$, as the curvature increases, photon counting statistics of the signal mode tends from the Poissonian to the sub-Poissonian, and for idler mode from sub-Poissonian to Poissonian. On the other hand, on a surface with a constant curvature, by increasing the value of $r$ statistics will change toward Poissonian, for both signal and idler modes. In addition, for a fixed $z$ and choosing $r = 0.5$, as the curvature increases, the behavior tends to sub-Poissonian for signal mode, and to Poissonian one for idler. All aforementioned analysis are in agreement with the the results of Fig. \ref{Fig:5}.

\begin{figure}[]
	\centering
	\subfloat[\centering ]{{\includegraphics[width=6.25cm]{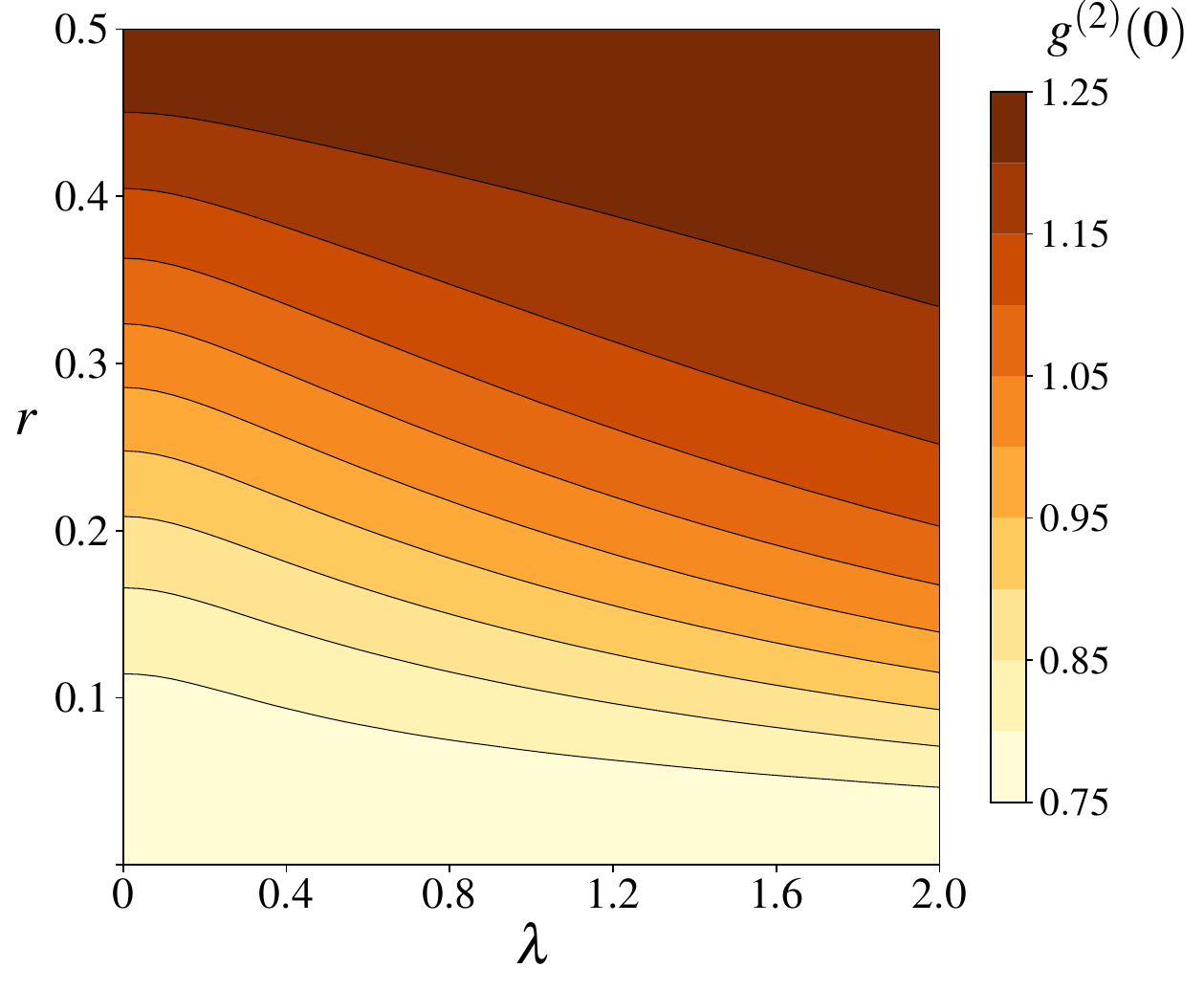}}}
	\label{fig:sub7.1}
	\qquad
	\subfloat[\centering ]{{\includegraphics[width=6.25cm]{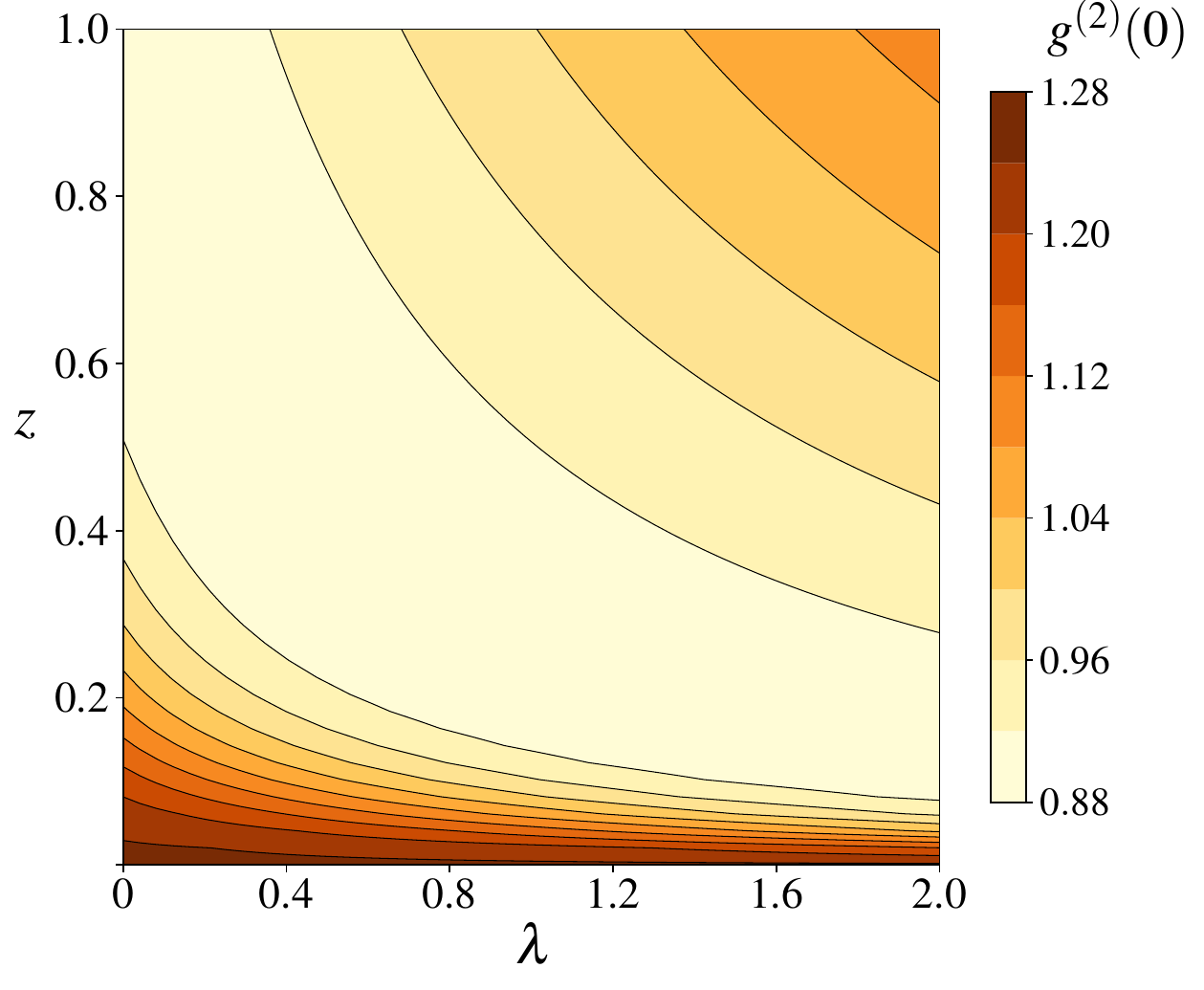}}}
	\label{fig:sub7.2}
	\caption{Contour plots of $g^{(2)}(0)$ are presented with $M = 4$. Contour (a) corresponds to $z = 1$, while (b) corresponds to $r = 0.5$}
	\label{Fig:7}
\end{figure}

\subsection{Normalized cross-correlation function}
Another useful parameter is the normalized cross-correlation function \cite{Ref34}:
\begin{equation}
	g^{(2)}(0) \equiv \frac{ \langle n_{s},n_{i}\rangle}{\langle n_{s} \rangle \langle n_{i} \rangle}.
	\label{Eq:23}
\end{equation}
This quantity describes correlation ($g^{(2)}(0) > 1$) or anti-correlation ($g^{(2)}(0) < 1$) between two modes. Fig.(\ref{Fig:7}) shows the contour plots of $g^{(2)}(0)$ with $M = 4$. Contour (a) corresponds to $z = 1$ and (b) corresponds to $r = 0.5$. It can be seen that, for a fixed value of $\lambda$ with $z = 1$, at $r = 0$ we have anti-correlated output modes, and by increasing $r$, they tend to become correlated. In addition, if we fix the $r$ parameter at $0.5$, two modes are correlated at $z=0$, but as this parameter increases, they tend to become anti-correlated.

\section{Summary and Concluding Remarks}
In this paper, by using an analog model of the general relativity, we have introduced a physical scheme that allows one to consider the curvature effects on the entanglement and quantum statistical properties of the Stim.PDC process. In fact, we have investigated in detail the entanglement generated via this process, when a two-mode SCSs served as two curvature-dependent input beams. We have used the linear entropy to measure the level of entanglement and studied the role of the curvature on it. It has been shown that the entanglement of the output states of curvature-dependent Stim.PDC is changing by the spatial curvature $\lambda$. As a consequence, the degree of entanglement can be controlled by seeding appropriate two-mode SCSs and using the suitable Stim.PDC parameters. Furthermore, we have considered the non-classical behaviors of the Stim.PDC output states, by using the Mandel parameter and the normalized cross-correlation function. to classical occurs at higher temperature.

\bibliographystyle{quantum}
\bibliography{Curvature}
\end{document}